\documentclass[adp,fleqn]{w-art}
\usepackage{times}
\usepackage{w-thm}
\usepackage[]{graphicx}
\chardef\bslash=`\\ 

\begin{document}
\DOIsuffix{theDOIsuffix}
\Volume{12}
\Issue{1}
\Copyrightissue{01}
\Month{01}
\Year{2003}
\pagespan{1}{}
\Receiveddate{}
\Accepteddate{}



\title[Single spin asymmetries]{Recent progress in the understanding 
of single spin asymmetries}


\author[A.\ Metz]{Andreas Metz\footnote{
     E-mail: {\sf metza@tp2.ruhr-uni-bochum.de}}} 
\address[]{Institut f\"ur Theoretische Physik II, 
 Ruhr-Universit\"at Bochum, D-44780 Bochum, Germany}
\author[M.\ Schlegel]{Marc Schlegel\footnote{
     E-mail: {\sf marcs@tp2.ruhr-uni-bochum.de}}} 
\dedicatory{Dedicated to Klaus Goeke on the occasion of his 
            $60^{th}$ birthday.}
\begin{abstract}
Over the past years a lot of progress has been made in the understanding
of single spin asymmetries in hard scattering processes.
We briefly review this subject, covering the non-vanishing of 
time-reversal odd parton distributions, universality of fragmentation 
functions, and the discovery of previously unknown time-reversal odd
parton distributions at subleading twist.
Interestingly enough, all these developments were initiated by simple 
model calculations.
\end{abstract}
\maketitle                   






\section{Introduction}
\label{sect1}

Non-zero single spin asymmetries (SSA) in hard scattering processes represent 
a very interesting phenomenon in particle physics.
For instance, large asymmetries (up to $40\%$) have been observed more than one 
decade ago in proton-proton collisions ($p\!\!\uparrow \!p \to \pi X$), where 
one of the protons is transversely polarized~\cite{E704_91}.
(For related theoretical work 
see~\cite{efremov_85,qiu_91,qiu_99,ratcliffe_99} as well as~\cite{anselmino_04} 
and references therein.)
Moreover, in lepton scattering off the nucleon non-zero results for various SSA 
have been reported recently~\cite{HERMES_00,CLAS_04}.
These results are surprising because in the leading twist collinear 
factorization approach for hard processes SSA strictly vanish~\cite{kane_78}.
In order to get non-zero effects one has to go beyond the collinear picture
and to take the transverse momentum of partons inside the hadrons into account.
In this context, a special role is played by transverse momentum dependent (TMD), 
time-reversal odd (T-odd) correlation functions (parton distributions and 
fragmentation functions) which are directly connected to the SSA.
Therefore, the recent progress in the understanding of SSA that has been achieved
over the past three years is actually progress in the understanding of the origin 
and of the properties of T-odd correlation functions.

The starting point was a model calculation for inclusive deep-inelastic 
scattering (DIS) which has shown the importance of gluon exchange between the 
struck quark and the target spectators~\cite{brodsky_02a}. 
It has been demonstrated that this rescattering effect causes (additional) 
on-shell intermediate states in the Compton amplitude, resulting in a 
modification of the DIS cross section.
In Feynman gauge, this (shadowing) effect is described by the gauge 
link appearing in the operator definiton of parton distributions.

Subsequently, the effect of rescattering has also been investigated in 
the case of semi-inclusive DIS~\cite{brodsky_02b}.
Using a simple spectator model, it has been shown that a non-zero SSA for
a transversely polarized target arises from the interference between the
tree-level amplitude of the fragmentation process and the imaginary
part of the one-loop amplitude, where the latter describes the gluon 
exchange between the struck quark and the target system.
This asymmetry has been interpreted as model for the T-odd Sivers 
parton distribution~\cite{sivers_90} including its gauge 
link~\cite{collins_02}.
The model calculation of Ref.~\cite{brodsky_02b} has therefore shown
for the first time explicitly that T-odd parton distributions can be 
non-zero, which is in contrast to the long-standing believe that they 
should vanish because of T-invariance of the strong 
interaction~\cite{collins_93}.
Consequently, T-odd parton distributions may (partly) be at the origin of 
the experimentally observed SSA.

While T-odd parton distributions are non-zero only if the gauge link is 
taken into account, T-odd fragmentation functions may well exist even if 
the rescattering associated with their gauge link is neglected.
Final state interactions can generate finite SSA in parton 
fragmentation~\cite{collins_93,collins_94,bacchetta_01}.
But the presence of the gauge link endangers universality of TMD fragmentation 
functions, i.e., it is a non-trivial question whether the fragmentation 
functions in DIS and in electron-positron annihilation coincide.
Whereas time-reversal can be used to relate parton distributions between 
DIS and the Drell Yan process~\cite{collins_02}, such an argument cannot 
be applied for fragmentation functions~\cite{boer_03,collins_04}.
However, a one-loop calculation for a SSA in fragmentation, analogous to 
the treatment of the target asymmetry in Ref.~\cite{brodsky_02b}, has 
shown universality for T-odd fragmentation functions~\cite{metz_02}.
In a recent work it has been argued that the finding in~\cite{metz_02}
is not a model-artefact but rather a general result~\cite{collins_04}.

In~\cite{afanasev_03,metz_04a} the spectator model calculation for SSA in 
semi-inclusive DIS of Ref.~\cite{brodsky_02b} has been extended to the case
of a longitudinally polarized lepton beam and a longitudinally polarized 
target.
For both asymmetries, which are of subleading twist (twist-3), non-zero
results have been found.
The results have indicated the existence of previously unknown subleading 
twist T-odd parton distributions~\cite{metz_04a}.
This conjecture has been confirmed recently by a model-independent 
analysis~\cite{bacchetta_04}. 

The purpose of this short review is to discuss the three mentioned developments
on the field of single spin asymmetries and TMD correlation functions, i.e., 
(1) the non-vanishing of T-odd parton distributions, (2) the universality of 
TMD fragmentation functions, (3) the discovery of so far unknown T-odd parton 
distributions at subleading twist.
Our emphasis will be on the model calculations which in all three cases preceded
the model-independent studies.

\section{Gauge invariant correlation functions}
\label{sect2}

In this section we summarize certain features of parton distributions and
fragmentation functions which are needed for the discussion lateron.
The classical process by which parton distributions of the nucleon are
measured is lepton-induced inclusive DIS, like $e^- p \to e^- X$.
This reaction provides information about the leading twist parton 
distributions 
$f_1$ (distribution of unpolarized quarks in an unpolarized nucleon) and 
$g_1$ (distribution of  longitudinally polarized quarks in a longitudinally
polarized nucleon). 
The third twist-2 distribution, the transversity $h_1$ (distribution of 
transversely polarized quarks in a transversely polarized nucleon), cannot 
be extracted from inclusive DIS data because of its chiral-odd 
nature~\cite{ralston_79}. 

A proper definition of these three distributions is given in terms of the
light-cone correlator
\begin{equation} \label{e:corr_1}
\Phi_{ij}(x,S)  =
\int \frac{d \xi^-}{2 \pi} \, e^{i x P^+ \xi^-} \,
 \langle P,S \, | \, \bar{\psi}_j(0) \,
 {\cal W}(0 , \xi^-) \,
 \psi_i(\xi^-) \, | \, P,S \rangle \, .
\end{equation}
In Eq.~(\ref{e:corr_1}) the target state is characterized by its four-momentum 
$P$ and the covariant spin vector $S$ 
$(P^2 = M^2, \; S^2 = -1 , \; P \! \cdot \! S = 0)$.
The longitudinal quark momentum is specified by $x$ via $k^+ = x P^+$.
The quark fields carry a Dirac index, while flavor and color indices 
are suppressed.
In order to ensure color gauge invariance of the correlator the Wilson-line 
\begin{equation}
{\cal W}(0,\xi^-) = {\cal P} \, {\text exp} 
 \bigg[ - i g \int_{0}^{\xi^-} d\eta^- A^+ (0,\eta^-,\vec{0}_{\perp}) \bigg] 
\end{equation}
connecting the two quark fields is needed.
It is well known that this gauge link encodes the re-interaction of the 
struck quark with the target system via the exchange of collinear gluons. 
In light-cone gauge ($A^+ = 0$) the link disappears.
The parton distributions can now be obtained from Eq.~(\ref{e:corr_1}) 
using suitable projections.
For instance, the unpolarized quark distribution is given by
$f_1(x) = {\rm Tr} (\Phi \gamma^+)/2$.

In inclusive DIS only the longitudinal momentum fraction of the struck 
parton is fixed by external kinematics, while its transverse momentum 
cannot be measured but is rather integrated over.
This situation changes in semi-inclusive DIS (like $e^- p \to e^- H X$), 
where in addition to the scattered electron a hadron $H$ is detected in 
the final state.
If the cross section is kept differential in the transverse momentum 
$\vec{P}_{h\perp}$ of the hadron, one is sensitive to both the 
transverse momentum on the parton distribution side (transverse momentum 
of the quark relative to the target) and on the fragmentation side. 
At tree level, it is relatively easy to establish factorization into TMD
parton distributions, TMD fragmentation functions and a hard 
scattering~\cite{ralston_79}.
The Drell-Yan process (like $p p \to \mu^+ \mu^- X$) with a low 
transverse momentum of the lepton-pair, and the reaction 
$e^+ e^- \to H_1 H_2 X$, where the two detected hadrons in back-to-back
jets have a low transverse momentum relative to each other, can be described 
within the same formalism of Ref.~\cite{ralston_79}.
Once gluonic corrections are included it becomes quite complicated to
establish QCD-factorization for this class of 
processes~\cite{collins_81,ji_04a,ji_04b}.
One source of the complications is the presence of soft gluons, which 
lead to an additional non-perturbative factor (soft factor) in a factorization 
formula. 
In the case of the one-loop calculations for T-odd correlation functions
to be discussed below these difficulties don't show up yet.
Therefore, we refrain from elaborating more on this issue here.

The correlator through which the TMD parton distributions are defined 
reads
\begin{equation} \label{e:corr_2}
\Phi_{ij}(x,\vec{k}_{\perp},S)  =
\int \frac{d \xi^- \, d^2 \vec{\xi}_{\perp}}{(2 \pi)^3} \,
 e^{i k \cdot \xi} \,
 \langle P,S \, | \, \bar{\psi}_j(0) \,
 {\cal W}(0 , \xi) \,
 \psi_i(\xi) \, | \, P,S \rangle \biggr|_{\xi^+=0} \, ,
\end{equation}
where the parton distributions can again be obtained using suitable
projections.
It turns out that eight TMD parton distributions can appear at leading 
twist~\cite{mulders_96,boer_98}, i.e., at leading order of a 
$1/Q$-expansion of observables, where $Q$ denotes the hard external 
momentum of the process.
In contrast to the correlation functions which show up in inclusive
processes, with two quark fields separated only along one light-cone 
direction, the quark fields in Eq.~(\ref{e:corr_2}) have also a separation 
in transverse position space.
Moreover, for a proper definition of TMD parton distributions the gauge 
link in (\ref{e:corr_2}) must not connect the quark fields by the shortest 
line but rather has to take the 
form~\cite{collins_82,collins_02,ji_02,belitsky_03}
\begin{equation} \label{e:lineDIS} 
 {\cal W}(0,\xi)_{DIS} =
 [0,0,\vec{0}_{\perp} ; 0,\infty,\vec{0}_{\perp}]
 \mbox{} \times
 [0,\infty,\vec{0}_{\perp};0,\infty,\vec{\xi}_{\perp}]
 \mbox{} \times
 [0,\infty,\vec{\xi}_{\perp};0,\xi^-,\vec{\xi}_{\perp}] \,.
\end{equation}
In this equation, $[a^+,a^-,\vec{a}_\perp;b^+,b^-,\vec{b}_\perp]$ denotes
the Wilson line connecting the points $a^\mu=(a^+,a^-,\vec{a}_\perp)$
and $b^\mu=(b^+,b^-,\vec{b}_\perp)$ along a straight line.
In covariant gauges like Feynman gauge the Wilson line at the light-cone 
infinity (second piece on the {\it rhs} in Eq.~(\ref{e:lineDIS})) 
can be neglected.
However, this line has to be taken into account in light-cone 
gauge~\cite{ji_02,belitsky_03}, because the transverse gluon potential 
doesn't vanish at the light-cone infinity.

It is now important to note that the path of the gauge link depends on 
the process.
While for DIS the Wilson lines are future-pointing, they are
past-pointing for Drell-Yan~\cite{collins_02,belitsky_03,boer_03},
\begin{equation} \label{e:lineDY} 
 {\cal W}(0,\xi)_{DY} =
 [0,0,\vec{0}_{\perp} ; 0,-\infty,\vec{0}_{\perp}]
 \mbox{} \times
 [0,-\infty,\vec{0}_{\perp};0,-\infty,\vec{\xi}_{\perp}]
 \mbox{} \times
 [0,-\infty,\vec{\xi}_{\perp};0,\xi^-,\vec{\xi}_{\perp}] \,.
\end{equation}
As a consequence, the definitions of TMD parton distributions are different 
for both processes, i.e., there is a danger of non-universality for these 
objects.
Exactly the same feature appears in the case of TMD fragmentation functions, 
which {\it a priori} have a different gauge link in semi-inclusive DIS 
compared to $e^+ e^-$ annihilation~\cite{boer_03}.
Whether and how one can still obtain universality for these TMD correlation
functions will be discussed in the following two sections.

Two out of the eight leading twist TMD parton distributions are T-odd, and
will be of particular importance for the subsequent discussion.
The same holds for T-odd fragmentation functions.
The following list (in the notation of Refs.~\cite{mulders_96,boer_98}) 
summarizes these functions and their meaning:
\begin{itemize}
\item $f_{1T}^{\perp}$: distribution of an unpolarized quark in a
 transversely polarized target~\cite{sivers_90}
\item $h_{1}^{\perp}$: distribution of a transversely polarized quark 
 in an unpolarized target~\cite{boer_98}
\item $D_{1T}^{\perp}$: fragmentation of an unpolarized quark into
 a transversely polarized hadron~\cite{mulders_96}
\item $H_{1}^{\perp}$: fragmentation of a transversely polarized quark 
 into an unpolarized hadron~\cite{collins_93}
\end{itemize}

T-odd correlation functions typically generate azimuthal/single spin 
asymmetries.
For instance, in the case of the Sivers function $f_{1T}^{\perp}$ one 
is dealing with a (T-odd) correlation 
$\vec{S}_{\perp} \cdot (\vec{P} \times \vec{k}_{\perp})$ between the
transverse target spin $\vec{S}_{\perp}$, the target momentum $\vec{P}$,
and the transverse momentum of the quark $\vec{k}_{\perp}$.
For a fixed target spin, this correlation gives rise to an azimuthal 
asymmetry of the quark distribution about the axis defined by the target 
momentum.
In 1993, it was proved that T-odd parton distributions should vanish 
because of T-invariance of the strong interaction~\cite{collins_93}.
(This claim lateron needed to be revised as will be explained in 
Sect.~\ref{sect3}.)
In contrast, T-odd fragmentation functions can exist because of final 
state interactions in the fragmentation process.
Such interactions can lead to a non-trivial phase (imaginary part) in the
scattering amplitude, which is needed to obtain a finite T-odd 
asymmetry~\cite{collins_93}.

\section{Transverse single spin asymmetry in DIS}
\label{sect3}

We now turn the attention to possible observable effects of the Wilson 
line discussed in the previous section, i.e., effects due to the rescattering 
of the struck quark via the exchange of longitudinally polarized gluons.
It has been demonstrated that even for inclusive DIS this rescattering 
influences the cross section~\cite{brodsky_02a}.
To be specific, it yields a reduction of the cross section which can be 
interpreted as shadowing effect~\cite{brodsky_02a}.

An observable being entirely connected with the gauge link has then
been considered in Ref.~\cite{brodsky_02b}.
In this work a simple model for a SSA with a transversely polarized 
spin-$\frac{1}{2}$ target has been constructed which will be discussed 
in some detail in this section.
The relevant process is
\begin{equation} \label{e:proc_DIS}
\gamma^{\ast}(q) + p(p,\lambda) \to q(p_1,\lambda') + s(p_2) \,,
\end{equation}
and is shown in Fig.~\ref{fig:1}.
The quark $q$ in the final state may either fragment into hadrons plus 
remnants or form a jet.
For our purpose, however, it is not necessary to consider the hadronization
of the quark.
A simple spectator model with a scalar diquark spectator $s$ is used for
the (proton) target~\cite{brodsky_02b}.
In this model the proton has no electromagnetic charge, and a charge $e_1$ 
is assigned to the quark. 
The interaction between the proton, the quark and the spectator is given
by a scalar vertex with the coupling constant $g$.

We treat the process~(\ref{e:proc_DIS}) in the Breit frame of the virtual 
photon. 
(Note that in~\cite{brodsky_02b} a different reference frame was used.)
The proton has a large plus-momentum $Q/x$, where 
$x = x_{Bj} + {\cal O}(1/Q^2)$.
The quark carries the large minus-momentum $p_1^- \approx q^-$ and a soft 
transverse momentum $\vec{\Delta}_{\perp}$.
These requirements specify the kinematics:
\begin{eqnarray} \label{e:kin}
& & q = \Big( -Q, \, Q, \, \vec{0}_{\perp} \Big) , \qquad
    p = \bigg( \frac{Q}{x}, \, \frac{xM^2}{Q}, \, \vec{0}_{\perp} \bigg) ,
\\
& & p_1 = \bigg( \frac{\vec{\Delta}_{\perp}^2 + m_q^2}{Q}, \, Q, \,
                 \vec{\Delta}_{\perp} \bigg) , \qquad
    p_2 = \bigg( \frac{Q (1-x)}{x}, \, 
                 \frac{x (\vec{\Delta}_{\perp}^2 + m_s^2)}{Q (1-x)}, \,
                -\vec{\Delta}_{\perp} \bigg) \,.
\nonumber
\end{eqnarray}
The expressions for $q$ and $p$ are exact, while for $p_1$ and $p_2$ just the
leading terms have been listed. 
In particular, sometimes the $1/Q^2$ corrections of $p_1^-$ and $p_2^+$ are 
needed which can be readily obtained from 4-momentum conservation.

\begin{vchfigure}[t!]
  \includegraphics[width=.7\textwidth]{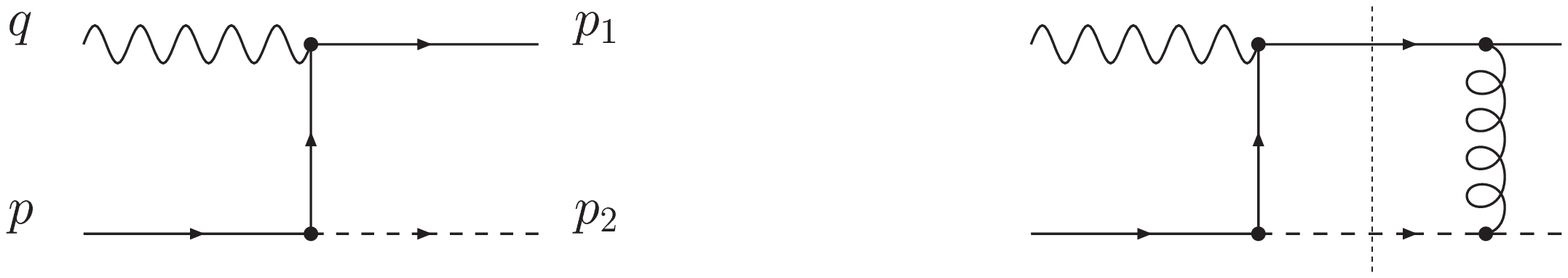}
\vchcaption{Tree-level and relevant one-loop contribution for single 
spin asymmetry in DIS in spectator model (see also text). The spectator
is indicated by a dashed line. The thin line characterizes the possible
on-shell intermediate state.}
\label{fig:1}
\end{vchfigure}

In the Breit frame the leading twist contribution is given by the two 
transverse components $J^{1,2}$ of the electromagnetic current.
(We define the current through the scattering amplitude via
$T = \varepsilon_{\mu} J^{\mu}$, with $\varepsilon$ denoting the polarization
vector of the virtual photon.) 
For our purpose it is sufficient to treat $J^1$, which reads for the
tree-level diagram on the {\it lhs} in Fig.~\ref{fig:1}
\begin{eqnarray} \label{e:tree}
J_{(0)}^{1}(\lambda,\lambda') & = &
 e_1 g \, \frac{1}{(p_1 - q)^2 - m_q^2} \, 
 \bar{u}(p_1,\lambda') \, \gamma^{1} \,
 (p_1 \! \cdot \! \gamma - q \! \cdot \! \gamma + m_q) \, u(p,\lambda) \,
\\
& = & - e_1 g \, \frac{1-x}{\sqrt{x} \, |\vec{\Delta}_{\perp}|} \,
 \frac{Q}{\vec{\Delta}_{\perp}^2 + \tilde{m}^2} 
 \bigg[ \Big( \lambda \, m_q ( Mx + m_q ) 
               -\lambda (\Delta^1 + i\lambda\Delta^2 )^2 \Big) 
        \delta_{\lambda,-\lambda'}
\nonumber \\
& & \hspace{2cm} + \Big( ( Mx + m_q ) ( \Delta^1 - i\lambda \Delta^2 ) 
             + m_q (\Delta^1 + i\lambda \Delta^2 ) \Big) 
        \delta_{\lambda,\lambda'} \bigg] ,
\nonumber \\
& & \textrm{with} \quad \tilde{m}^2 = x (1 - x) 
                 \bigg( - M^2 + \frac{m_q^2}{x} + \frac{m_s^2}{1 - x} \bigg) .
\nonumber
\end{eqnarray}

The transverse SSA is given by the ratio $\sigma_{pol} / \sigma_{unp}$, where 
the unpolarized and the polarized (polarization along the $x$-axis) cross 
sections are computed according to
\begin{eqnarray} 
\sigma_{unp} & \propto & \frac{1}{2} \sum_{\lambda,\lambda'}
 \Big( J^1 (\lambda,\lambda') \Big)^{\dagger} J^1 (\lambda,\lambda') \,,
\\  
\sigma_{pol} & \propto & \frac{1}{2} \sum_{\lambda'}
 \bigg[ \Big( J^1 (s_x = \uparrow,\lambda') \Big)^{\dagger} 
              J^1 (s_x = \uparrow,\lambda')
      - \Big( J^1 (s_x = \downarrow,\lambda') \Big)^{\dagger} 
              J^1 (s_x = \downarrow,\lambda') \bigg] \,.
\end{eqnarray}
A tree-level calculation taking only the expression in Eq.~(\ref{e:tree})
into account leads to a vanishing asymmetry because the scattering amplitude 
has no imaginary part.
In order to generate a finite imaginary part loop corrections have to be 
included, where we limit the analysis to the one-loop approximation. 
For simplicity, the gluon exchange is modelled by an Abelian gauge 
field~\cite{brodsky_02b}.
It is easy to see that for the process in~(\ref{e:proc_DIS}) only the box 
graph on the {\it rhs} in Fig.~\ref{fig:1}, which describes the re-interaction 
of the struck quark with the target system to lowest order, gives rise 
to an imaginary part. 
Now one indeed obtains a non-zero transverse SSA which is given by the 
interference of the tree-level amplitude with the imaginary part of the 
one-loop amplitude.
Explicitly, one finds~\cite{brodsky_02b}
\begin{equation} \label{e:asy}
A_{UT,x} = - \frac{(e_1)^2}{8 \pi} \, \frac{2 (Mx + m_q) \Delta^2}
                             {(Mx + m_q)^2 + \vec{\Delta}_{\perp}^2} \,
            \frac{\vec{\Delta}_{\perp}^2 + \tilde{m}^2}{\vec{\Delta}_{\perp}^2} \,
            \ln \frac{\vec{\Delta}_{\perp}^2 + \tilde{m}^2}{\tilde{m}^2} \,.
\end{equation}
Obviously, the asymmetry vanishes if the transverse momentum of the quark 
vanishes.
We emphasize again that the SSA would be zero if there were no 
re-interaction of the struck quark.

Soon after the publication of Ref.~\cite{brodsky_02b} it has been realized 
that the asymmetry in~(\ref{e:asy}) is actually not a new effect, but can 
rather be understood as a model calculation for the T-odd Sivers function
including its gauge link~\cite{collins_02}.
Therefore, the work in~\cite{brodsky_02b} has demonstrated for the first
time explicitly that T-odd parton distributions can be non-zero.
This result was at variance with the proof~\cite{collins_93} according to 
which T-odd parton distributions should vanish because of T-invarince of the
strong interaction.
In fact it turned out that the proof no longer holds once the gauge link 
is taken into account~\cite{collins_02}.

\begin{vchfigure}[t!]
  \includegraphics[width=.7\textwidth]{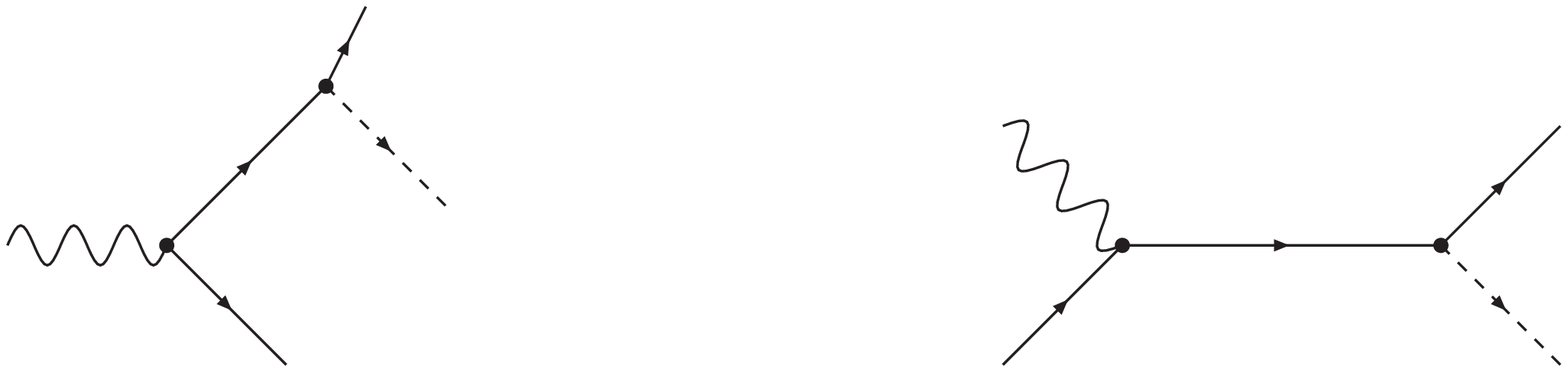}
\vchcaption{Tree-level diagrams for fragmentation in $e^+ e^-$ annihilation
and semi-inclusive DIS. In both cases a quark fragments into a 
spin-$\frac{1}{2}$ hadron and a scalar remnant (dashed line).}
\label{fig:2}
\end{vchfigure}

As mentioned in Sect.~\ref{sect2}, the parton distributions in DIS and 
Drell-Yan have a different gauge link which endangers their universality.
Nevertheless, the two definitions can still be related using 
time-reversal. 
It has been found that all six T-even TMD parton distributions are 
universal, while a violation of universality appears only for the two 
T-odd functions in the sense that they have a reversed sign in both 
processes~\cite{collins_02},
\begin{eqnarray} \label{e:univpdf_a}
f_{1T}^{\perp} \Big |_{DIS} & = & - f_{1T}^{\perp} \Big |_{DY} \,,
 \\  \label{e:univpdf_b}
h_{1}^{\perp} \Big |_{DIS} & = & - h_{1}^{\perp} \Big |_{DY} \,.
\end{eqnarray}
From a practical point of view this violation of universality has no
direct consequence, since the possibility of relating cross sections 
of different processes is not spoiled.
An experimental investigation of the 
relations~(\ref{e:univpdf_a},\ref{e:univpdf_b}) would serve as a very 
important check of our present-day understanding of the nature 
of T-odd parton distributions and, hence, our understanding of SSA.

\section{Universality of fragmentation functions}
\label{sect4}

Now we want to investigate the influence of the Wilson line on the 
fragmentation process.
Fragmentation functions in DIS and $e^+ e^-$ annihilation {\it a priori} 
have different Wilson lines (future-pointing for $e^+ e^-$, 
past-pointing for DIS).
However, in contrast to parton distributions, they cannot be related using 
time-reversal because they involve a semi-inclusive sum over 
out-states~\cite{boer_03,collins_04},
\begin{equation}
\sum_{X} \, |A, X, {\rm out} \rangle \langle A, X, {\rm out} | \,.
\end{equation}
Time-reversal converts these into in-states, and therefore is useless in
order to establish any relation between the fragmentation functions for 
both reactions. 
This means that fragmentation functions are non-universal, unless another 
argument comes to our aid.
To investigate this point a model calculation for a transverse SSA in 
fragmentation (unpolarized quark into transversely polarized hadron, 
i.e., a model for $D_{1T}^{\perp}$) has been performed~\cite{metz_02}, 
which is quite similar to the calculation discussed in the previous 
section.

In Fig.~\ref{fig:2}, the tree-level diagrams of the fragmentation in 
$e^+ e^-$ annihilation and in DIS are displayed.
For $e^+ e^-$ annihilation we consider the decay of a timelike virtual
photon into a $q\bar{q}$ pair, where the quark fragments into a 
spin-$\frac{1}{2}$ hadron (e.g. a proton) and a scalar remnant, i.e.,
\begin{equation}
 \gamma^{\ast} \to \bar{q} + p + s \,. 
\end{equation}
The fragmentation of the quark is described in the model of 
Ref.~\cite{brodsky_02b} used in Sect.~\ref{sect3}.
The one-loop corrections are shown in Fig.~\ref{fig:3}. 
For $e^+ e^-$ annihilation (semi-inclusive DIS) a single Abelian gluon 
is exchanged between the remnant and the antiquark (initial quark).
These diagrams provide a simple model for the lowest order contribution 
of the gauge link of the fragmentation function.
Two cuts (on-shell quark and antiquark, as well as on-shell
antiquark and remnant) for $e^+ e^-$ annihilation have no
counterpart in semi-inclusive DIS.
The quark-photon cut in $e^+ e^-$ annihilation corresponds to the 
cut in DIS.
Note that the one-loop correction at the photon vertex, which is also 
related to the gauge link of the fragmentation function, is irrelevant
for a transverse SSA, i.e., it neither contributes to $D_{1T}^{\perp}$
nor to $H_{1}^{\perp}$.
All remaining one-loop graphs have no relation to the gauge link.
These effects are obviously universal (see, e.g., also Ref.~\cite{teryaev_03}),
and won't be considered in the following.

The calculation for the transverse SSA is similar to the case of the 
target asymmetry considered in Sect.~\ref{sect3}.
Here we just focus on the main results, while detailed formulae can be 
found in Ref.~\cite{metz_02}. 
We first consider the two on-shell intermediate states in 
$e^+ e^-$ annihilation which have no counterpart in DIS and, hence, 
form a potential source of non-universality of the transverse SSA.
However, it turns out that the contributions of both intermediate 
states to the SSA cancel each other~\cite{metz_02}, 
\begin{equation} \label{e:cancel}
{\cal A}^{\bar{q} q} \Big|_{e^+ e^-} = -
{\cal A}^{\bar{q} s} \Big|_{e^+ e^-}  \,.
\end{equation}
This cancellation is exact for the leading twist term of the asymmetry,
which is sufficient for our investigation here.
At subleading twist the equality~(\ref{e:cancel}) breaks down.

One is now left with the on-shell $q g$ intermediate state only.
By explicit calculation it has been shown that the corresponding
contribution to the SSA coincides for both processes~\cite{metz_02},
\begin{equation}
{\cal A}^{q g} \Big|_{DIS} = 
{\cal A}^{q g} \Big|_{e^+ e^-} \,,
\end{equation}
i.e., the asymmetry in both processes has the same sign.
In particular, one does not observe a sign-reversal for the one-loop box 
graph that is associated with the gauge link of the fragmentation 
function.
The origin for the different behaviour compared to the T-odd target
asymmetry is directly related to the different kinematics one is 
dealing with in fragmentation.
Note for instance that, in contrast to the case of the target asymmetry, 
in fragmentation the surviving contribution to the SSA involves a cut of the 
gauge boson propagator.

\begin{vchfigure}[t!]
  \includegraphics[width=.7\textwidth]{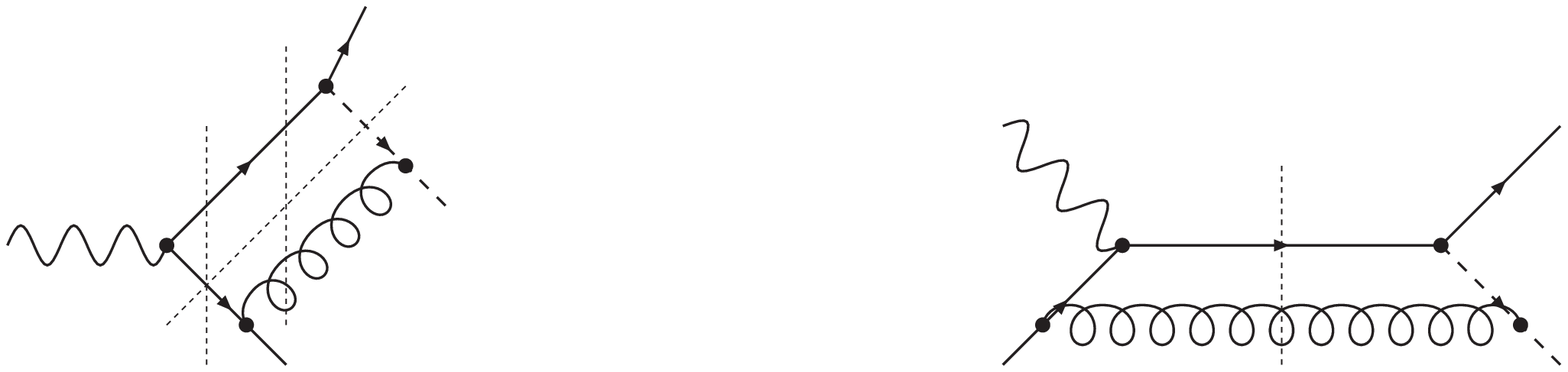}
\vchcaption{One-loop diagrams for fragmentation in $e^+ e^-$ annihilation
and semi-inclusive DIS. The possible on-shell intermediate states are
indicated by thin lines.}
\label{fig:3}
\end{vchfigure}

Altogether, one finds that the total transverse SSA in $e^+ e^-$ annihilation 
and in semi-inclusive DIS are equal, i.e, the T-odd fragmentation of an 
unpolarized quark into a transversely polarized spin-$\frac{1}{2}$ hadron 
is universal in the one-loop model.
This implies universality of the fragmentation function $D_{1T}^{\perp}$.
The same conclusion holds for the Collins function as well~\cite{metz_02}.
Therefore, we have the results
\begin{eqnarray}
D_{1T}^{\perp} \Big |_{DIS} & = & D_{1T}^{\perp} \Big |_{e^+ e^-} \,,
 \\
H_{1}^{\perp} \Big |_{DIS} & = & H_{1}^{\perp} \Big |_{e^+ e^-} \,.
\end{eqnarray}

This model calculation has indicated that fragmentation functions might
well be universal despite the {\it a priori} reversed Wilson lines they have 
in $e^+ e^-$ annihilation compared to semi-inclusive DIS.
Recently, it has indeed been shown by a more general analysis that this 
suspicion is true~\cite{collins_04}.
Universality not only holds for the two T-odd fragmentation functions
but also for the T-even ones.
Moreover, it has been argued that it should be valid to all orders in 
perturbation theory.
The crucial observation is that, due to the special analytical structure
of the fragmentation process, one is not sensitive to the direction of the 
Wilson lines.
In particular, factorization for semi-inclusive DIS can be derived with both
past-pointing and future-pointing Wilson lines for the fragmentation 
functions~\cite{collins_04}. 

\begin{vchfigure}[t!]
  \includegraphics[width=.7\textwidth]{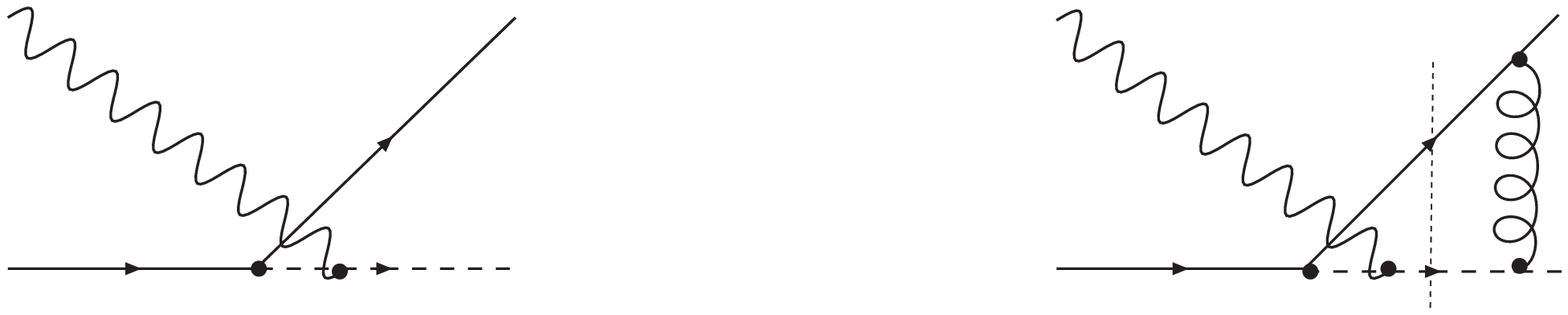}
\vchcaption{Tree-level and one-loop contribution to the process 
in~(\ref{e:proc_DIS}).
Such diagrams, where the photon couples to the spectator, are needed in order 
to ensure electromagnetic gauge invariance at subleading twist.}
\label{fig:4}
\end{vchfigure}

\section{Longitudinal single spin asymmetries in DIS}
\label{sect5}

The transverse SSA $A_{UT}$ in semi-inclusive DIS discussed in 
Sect.~\ref{sect3} is a leading twist observable.
One may wonder whether also the subleading asymmetries 
$A_{UL}$ (longitudinally polarized target) and 
$A_{LU}$ (longitudinally polarized lepton) are affected by the rescattering
of the struck quark.
It is quite important to clarify this issue, because most of the 
existing data and of the analyses 
(see e.g. Refs.~\cite{oganessyan_98,efremov_00,desanctis_00,ma_01,efremov_01,efremov_03,yuan_04,gamberg_04})
are concerned with these twist-3 observables.

In most of the theoretical papers a formalism is used which only includes T-odd 
fragmentation functions~\cite{levelt_94,mulders_96}.
In this scenario one finds schematically $A_{LU} \propto e \, H_1^{\perp}$ and
$A_{UL} \propto h_L \, H_1^{\perp}$, with the twist-3 T-even parton distributions
$e$ and $h_L$. 
The most recent works in Refs.~\cite{yuan_04,gamberg_04} on the beam spin
asymmetry extended the analysis by considering also T-odd parton distributions.
This leads for $A_{LU}$ to an additional term $h_1^{\perp} \, E$, where $E$ is a 
twist-3 fragmentation function.
However, it was still unclear if also possible higher twist T-odd distribution 
functions, generated by the same rescattering mechanism that leads to twist-2 
T-odd functions, could be relevant for the longitudinal SSA.

In order to proceed at this point, the spectator model calculation for $A_{UT}$ 
in Ref.~\cite{brodsky_02b} was extended to the twist-3 case~\cite{afanasev_03}.
In this work, $A_{LU}$ was computed from the Feynman diagrams in Fig.~\ref{fig:1}, 
and a non-zero result was obtained.
But it turned out that, at the twist-3 level, the graphs in Fig.~\ref{fig:1} 
no longer satisfy electromagnetic gauge invariance.
This has been corrected in Ref.~\cite{metz_04a} by including the diagrams
in Fig.~\ref{fig:4}, where the photon couples to the diquark spectator.
Moreover, also the target spin asymmetry $A_{UL}$ has been computed 
in~\cite{metz_04a}.
The gauge invariant treatment yields non-zero results for both asymmetries,  
\begin{equation}
A_{LU} \neq 0 \,, \qquad A_{UL} \neq 0 \,,
\end{equation}
where details of the calculation can be found in~\cite{metz_04a}.

Let us now turn to the implications of this calculation.
First of all, it is obvious that the diagrams in Fig.~\ref{fig:4} are not 
compatible with the parton model, and therefore are also not compatible with
factorization.
However, this does not necessarily mean that factorization for semi-inclusive
DIS is broken at subleading twist, even if it has not yet been proved.
One rather reaches the limitations of the perturbative spectator model for
the nucleon.
In particular, one finds that large momentum transfers at the 
proton-quark-diquark vertex, which appear in the graphs in Fig.~\ref{fig:4}, 
are not suppressed. 
Actually they get enhanced, which is at variance with the parton model.

Despite this shortcoming of the spectator model, it has been argued
that in a parton model description of $A_{LU}$ and $A_{LU}$ subleading 
T-odd distributions should be present~\cite{metz_04a}.
Mainly for two reasons this speculation has been made:
first, within the model calculation non-zero asymmetries arise already 
from the diagrams in Fig.~\ref{fig:1} whose kinematics is compatible 
with the parton model. 
Second, there is no reason why the asymmetries should not contain higher
order T-odd distribution functions.

A revised parton model analysis for $A_{LU}$ and $A_{UL}$ lateron
has confirmed this speculation~\cite{bacchetta_04}. 
Both asymmetries contain an additional term with a twist-3 T-odd distribution 
function, which have not been taken into account in the literature before.
The main observation has been that additional functions are generated once
the gauge link in the correlator in Eq.~(\ref{e:corr_2}) is taken into 
account~\cite{goeke_03,bacchetta_04}.

\section{Conclusions}
\label{sect6}

We have summarized important recent progress in the understanding
of single spin asymmetries in hard scattering processes.
All these developments are based on the fact that collinear gluon exchange 
(e.g., in DIS between the struck quark and the target system) leads to 
observable effects.
In a factorized description these contributions are encoded in the gauge
link of parton distributions and fragmentation functions.

The most striking influence of collinear gluons appears in the case
of single spin asymmetries/T-odd correlation functions.
While T-odd fragmentation functions are non-zero without such gluon
exchange, T-odd parton distributions vanish if their gauge link is 
neglected.
The existence of T-odd parton distributions considerably enriches the 
phenomenology of single spin asymmetries.
For instance, in semi-inclusive DIS such effects not only enter the leading
twist asymmetry $A_{UT}$, but also the twist-3 observables 
$A_{UL}$ and $A_{LU}$.
Since the gauge link is process dependent, it endangers the universality
of transverse momentum dependent correlation functions.
However, it has been shown that the only violation of universality 
appears for T-odd parton distributions in terms of a reversed sign 
between DIS and the Drell-Yan process.

Eventually, our short review nicely shows that model calculations quite 
often lead to the discovery of new model-independent results.
This remains true despite the limitations models typically have.

\begin{acknowledgement}
 We are very grateful to Klaus Goeke for the pleasant and stimulating 
 atmosphere at the {\it Institut fuer Theoretische Physik II, 
 Ruhr-Universit\"at Bochum.}
 One of us (A.M.) would like to thank John Collins for fruitful
 collaboration. 
 The work has been partly supported by the DFG, and the Sofia 
 Kovalevskaya Programme of the Alexander von Humboldt Foundation.
\end{acknowledgement}



\begin{thebibliography}{10}

\bibitem{E704_91}
D. L. Adams et al., Phys. Lett B \textbf{264}, 462 (1991).

\bibitem{efremov_85}
A. V. Efremov and O. V. Teryaev, 
Phys. Lett. B \textbf{150}, 383 (1985).

\bibitem{qiu_91}
J. Qiu and G. Sterman, Phys. Rev. Lett. \textbf{67}, 2264 (1991). 

\bibitem{qiu_99}
J. Qiu and G. Sterman, Phys. Rev. D \textbf{59}, 014004 (1999). 

\bibitem{ratcliffe_99}
P. G. Ratcliffe, Eur. Phys. J. \textbf{C8}, 403 (1999).

\bibitem{anselmino_04}
M. Anselmino, M. Boglione, U. D'Alesio, E. Leader, and F. Murgia,
hep-ph/0408356.

\bibitem{HERMES_00}
HERMES Collaboration, A. Airapetian, et al., 
Phys. Rev. Lett. \textbf{84}, 4047 (2000);
Phys. Rev. D \textbf{64}, 097101 (2001);
Phys. Lett. B \textbf{562}, 182 (2003);
hep-ex/0408013.

\bibitem{CLAS_04}
CLAS Collaboration, H. Avakian, et al., Phys. Rev. D \textbf{69}
112004 (2004).

\bibitem{kane_78}
G. L. Kane, J. Pumplin, and W. Repko, Phys. Rev. Lett. \textbf{41},
1689 (1978).

\bibitem{brodsky_02a}
S. J. Brodsky, P. Hoyer, N. Marchal, S. Peign\'e, and F. Sannino,
Phys. Rev. D \textbf{65}, 114025 (2002).

\bibitem{brodsky_02b}
S. J. Brodsky, D. S. Hwang, and I.~Schmidt, Phys. Lett. B \textbf{530},
99 (2002).

\bibitem{sivers_90}
D. W. Sivers, Phys. Rev. D \textbf{41}, 83 (1990); 
Phys. Rev. D \textbf{43}, 261 (1991).

\bibitem{collins_02}
J. C. Collins, Phys. Lett. B \textbf{536}, 43 (2002).

\bibitem{collins_93}
J. C. Collins, Nucl. Phys. \textbf{B396}, 161 (1993).

\bibitem{collins_94}
J. C. Collins and G. A. Ladinsky, hep-ph/9411444.

\bibitem{bacchetta_01}
A. Bacchetta, R. Kundu, A. Metz, and P. J. Mulders, Phys. Lett. B \textbf{506},
155 (2001).

\bibitem{boer_03}
D. Boer, P. J. Mulders, and F. Pijlman, Nucl. Phys. \textbf{B667}, 201 (2003).

\bibitem{collins_04}
J. C. Collins and A. Metz, hep-ph/0408249.

\bibitem{metz_02}
A. Metz, Phys. Lett. B \textbf{549}, 139 (2002).

\bibitem{afanasev_03}
A. Afanasev and C. E. Carlson, hep-ph/0308163.

\bibitem{metz_04a}
A. Metz and M. Schlegel, hep-ph/0403182, to appear in EPJA.

\bibitem{bacchetta_04}
A. Bacchetta, P. J. Mulders, and F. Pijlman, Phys. Lett. B \textbf{595},
309 (2004).

\bibitem{ralston_79}
J. P. Ralston and D. E. Soper, Nucl. Phys. \textbf{B152}, 109 (1979).

\bibitem{collins_81}
J. C. Collins and D. E. Soper, Nucl. Phys. \textbf{B193}, 381 (1981);
Nucl. Phys. \textbf{B213}, 545 (1983) (E).

\bibitem{ji_04a}
X. Ji, J. Ma, and F. Yuan, hep-ph/0404183.

\bibitem{ji_04b}
X. Ji, J. Ma, and F. Yuan, Phys. Lett. B \textbf{597}, 299 (2004).

\bibitem{mulders_96}
P. J. Mulders and R. D. Tangerman, Nucl. Phys. \textbf{B461}, 197 (1996);
Nucl. Phys. \textbf{B484}, 538 (1997) (E).

\bibitem{boer_98}
D. Boer and P. J. Mulders, Phys. Rev. D \textbf{57}, 5780 (1998).

\bibitem{collins_82}
J. C. Collins and D. E. Soper, Nucl. Phys. \textbf{B194}, 445 (1982).

\bibitem{ji_02}
X. Ji and F. Yuan, Phys. Lett. B \textbf{543}, 66 (2002).

\bibitem{belitsky_03}
A. V. Belitsky, X. Ji, and F. Yuan, Nucl. Phys \textbf{B656}, 165 (2003).

\bibitem{teryaev_03}
O. V. Teryaev, Czech. J. Phys. \textbf{53}, 47A (2003).

\bibitem{oganessyan_98}
K. A. Oganessyan, H. R. Avakian, N. Bianchi, and A. M. Kotzinian,
hep-ph/9808368.

\bibitem{efremov_00}
A. V. Efremov, K. Goeke, M. V. Polyakov, and D. Urbano, 
Phys. Lett. B \textbf{478}, 94 (2000).

\bibitem{desanctis_00}
E. De Sanctis, W. D. Nowak, and K. A. Oganessyan, 
Phys. Lett. B \textbf{483}, 69 (2000).

\bibitem{ma_01}
B. Q. Ma, I. Schmidt, and J. J. Yang, Phys. Rev. D \textbf{63}, 
037501 (2001).

\bibitem{efremov_01}
A. V. Efremov, K. Goeke, and P. Schweitzer, 
Phys. Lett. B \textbf{522}, 37 (2001);
Phys. Lett. B \textbf{544}, 389 (2002) (E).

\bibitem{efremov_03}
A. V. Efremov, K. Goeke, and P.~Schweitzer, Phys. Rev. D \textbf{67}, 
114014 (2003).

\bibitem{yuan_04}
F. Yuan, Phys. Lett. B \textbf{589}, 28 (2004).

\bibitem{gamberg_04}
L. P. Gamberg, D. S. Hwang, and K. A. Oganessyan, 
Phys. Lett. B \textbf{584}, 276 (2004). 

\bibitem{levelt_94}
J. Levelt and P. J. Mulders, Phys. Lett. B \textbf{338}, 357 (1994).

\bibitem{goeke_03}
K. Goeke, A. Metz, P. V. Pobylitsa, and M. V. Polyakov,
Phys. Lett. B \textbf{567}, 23 (2003).

\end{thebibliography}
\end{document}